\nonstopmode

\newcommand{\I}[1]{_{\mathrm{#1}}}
\newcommand{\Int}{\int\limits}
\newcommand{\differential}{\>\mathrm d}

\documentclass[a4paper]{jpconf}
\usepackage{graphicx}
\usepackage{bm,amsmath,amssymb,cite}

\begin{document}
\title{Strong-field control of x-ray absorption}
\author{R~Santra, C~Buth, E~R~Peterson, R~W~Dunford, E~P~Kanter,
B~Kr\"assig, S~H~Southworth and L~Young}
\address{Argonne National Laboratory, Argonne, Illinois 60439, USA}
\ead{rsantra@anl.gov}

\begin{abstract}
Strong optical laser fields modify the way x rays interact with matter.
This allows us to use x rays to gain deeper insight into strong-field
processes.  Alternatively, optical lasers may be utilized to control
the propagation of x rays through a medium.  Gas-phase systems are
particularly suitable for illustrating the basic principles underlying
combined x-ray and laser interactions.  Topics addressed
include the impact of spin-orbit interaction on the alignment of
atomic ions produced in a strong laser field, electromagnetically
induced transparency in the x-ray regime, and laser-induced alignment
of molecules.
\end{abstract}

\section{Introduction}
\label{sec1}

In x-ray science, cross sections for x-ray absorption in matter are
typically assumed to be invariants, depending only on the atomic
composition of the material \cite{Attw00,AlMc01}. However, by utilizing
electromagnetic fields, it is possible to modify x-ray absorption cross
sections in a controlled way. Wuilleumier and Meyer \cite{WuMe06} review
combined laser/x-ray experiments performed at laser intensities that are
small in comparison to an atomic unit ($\sim 10^{16}$~W/cm$^2$), so that
the laser--matter interaction may be described perturbatively. As
discussed in reference~\cite{Yama02}, at laser intensities above $\sim
10^{10}$~W/cm$^2$, interesting processes such as tunnel ionization
\cite{DeKr00} and molecular alignment \cite{StSe03} can occur. In the
following, we discuss three different scenarios for modifying x-ray
absorption near an inner-shell edge. Strong-field ionization at
$10^{14}$~W/cm$^2$ is the topic of section~\ref{sec2}. In
section~\ref{sec3}, laser dressing at $10^{13}$~W/cm$^2$ is discussed.
Section~\ref{sec4} presents results of a recent joint theoretical and
experimental study on x-ray absorption by laser-aligned molecules
($10^{12}$~W/cm$^2$). For the purpose of this paper, the x-ray regime is
defined by photon energies that are sufficient to produce, via
photoabsorption, inner-shell vacancies in atoms such as carbon or
heavier. This definition includes both soft \cite{Attw00} and hard
\cite{AlMc01} x rays. We concentrate on a laser wavelength of $800$~nm.
In this paper, the laser and x-ray fields are assumed to be linearly
polarized and co-propagating.

\section{X-ray probe of strong-field ionized atoms}
\label{sec2}

Reference \cite{YoAr06} develops an x-ray microprobe methodology that
allows one to study strong-field ionized atoms or molecules using x-ray
absorption spectroscopy. X rays are focused within the focal volume of a
strong optical field such that the x-ray focal width is much smaller (by
a factor of $10$ in reference~\cite{YoAr06}) than the laser focal width.
X-ray absorption is monitored by measuring the x-ray fluorescence
resulting from inner-shell hole formation within an interval along the
laser propagation axis that is short in comparison with the Rayleigh
range \cite{YoAr06}. Thus, the x-ray microprobe methodology reduces
averaging effects over the spatial intensity distribution of the laser
field. The species first studied in the strong-field experiments carried
out at Argonne's Advanced Photon Source was krypton
\cite{YoAr06,HoPe07}.

\begin{figure}[ht]
  \includegraphics[clip,width=25pc]{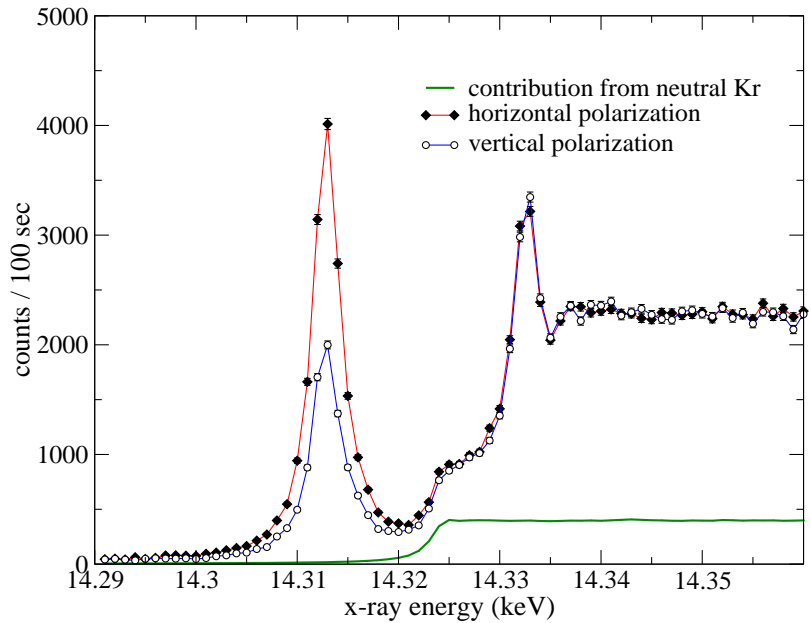}\hspace{2pc}%
  \begin{minipage}[b]{10pc}
    \caption{\label{fig1} Experimental near-{\em K}-edge absorption spectra of
         strong-field ionized Kr.
         Data are shown for both parallel and perpendicular laser and
         x-ray polarizations.
         Also depicted is the background contribution from neutral Kr.}
  \end{minipage}
\end{figure}

Figure~\ref{fig1} shows the Kr {\em K}$\alpha$ fluorescence signal from
krypton gas in a gas cell (number density $5\times 10^{15}$~cm$^{-3}$)
as a function of the energy of the absorbed x-ray photon. The signal was
measured $0.5$~ns after a $50$-fs, $\sim 10^{14}$-W/cm$^2$ laser pulse
ionized the Kr atoms. The full diamonds/red lines show the fluorescence
signal for laser polarization parallel to the x-ray polarization; the
open circles/blue lines refer to perpendicular polarization directions.
For the case shown, the majority of atoms are ionized. There are two
prominent features in the spectrum resulting from Kr$^+$ $1s \rightarrow
4p$ and $1s \rightarrow 5p$ excitations. The contribution of
fluorescence from neutral atoms in the interaction region is indicated
by the green line. The ratio of ions to neutrals is about $4.7:1$. It
may be concluded from the spectrum shown that there is very little
Kr$^{2+}$ present, because the well-separated Kr$^{2+}$ $1s \rightarrow
5p$ excitation line seen at higher laser intensities in reference
\cite{YoAr06} is hardly noticeable here.

There are two interesting observations we would like to point out.
First, strong-field ionization modifies the structure observed in the
near-edge absorption spectrum quite dramatically. For instance, when Kr
spectra are measured in the absence of the laser, there is very little
absorption at an x-ray photon energy of $14.313$~keV. In the presence of
the laser, there appears at this photon energy the strong Kr$^+$ $1s
\rightarrow 4p$ absorption resonance seen in figure~\ref{fig1}. (See
reference~\cite{HeAd06} for a similar observation made in potassium.)
The effect is very pronounced, but the disadvantage of this form of
strong-field control of x-ray absorption is that it is not rapidly
reversible: One has to wait until the laser-produced plasma has expanded
\cite{HoPe07} and neutral atoms have filled the original plasma volume.

The second observation is that the Kr$^+$ $1s \rightarrow 4p$ resonance
displays a strong dependence on the relative polarization of laser and x
rays. The resonant x-ray absorption cross section for parallel
polarizations is two times larger than in the perpendicular case. The
observed linear dichroism provides information on the alignment of the
$4p$ hole orbital produced by the strong laser pulse
\cite{YoAr06,SaDu06}. The hole orbital is aligned along the laser
polarization axis, which defines the quantization axis. As shown in
references~\cite{YoAr06,SaDu06}, in order to understand the
observations, it is necessary to take into account the effect of
spin-orbit coupling. Nonrelativistic tunneling or multiphoton ionization
models strongly overestimate the degree of ion alignment. According to
adiabatic strong-field ionization calculations that include spin-orbit
coupling \cite{SaDu06}, there is hardly any mixing between the
$4p_{3/2}$ and $4p_{1/2}$ orbitals at the electric-field strength at
which Kr ionizes. In other words, the laser field is not strong enough
to break spin-orbit coupling. The same is true in the case of
strong-field ionization of Xe \cite{LoKh07}. Therefore, if there are
coherences in the density matrix of Kr$^+$ or Xe$^+$, they are not
caused by the strength of the laser electric field.

\begin{table}
  \caption[]{\label{tab1} Kr$^+$ quantum state populations $\rho_{j,\vert m \vert}$,
         following strong-field ionization of Kr, obtained from experiment
         and theory.
         The errors shown reflect only statistical uncertainties ($2\sigma$).
         Potential systematic errors are not included.}
  \begin{center}
    \begin{tabular}{llll}
\br
                  & $\rho_{3/2,1/2}$~($\%$) & $\rho_{1/2,1/2}$~($\%$) & $\rho_{3/2,3/2}$~($\%$)\\
\mr
Experimental      & $59 \pm 6$              & $35 \pm 4$              & $6 \pm 6$\\
Theoretical       & $71$                    & $25$                    & $4$\\
\br
    \end{tabular}
  \end{center}
\end{table}

In reference \cite{LoKh07} it was demonstrated for Xe that inner-shell
absorption spectra of laser-produced ions may be used to determine the
quantum state populations $\rho_{j,\vert m \vert}$, i.e., the
probability to find an ion with a hole in the $np_j$ orbital with
projection quantum number $\pm m$ ($n$ is the principal quantum number
of the valence shell). In table~\ref{tab1}, we present the results of
such an analysis for Kr. The analysis is based on the fact that using
standard angular momentum algebra, it may be shown for the $1s
\rightarrow 4p$ transition in Kr$^+$ that the x-ray absorption cross
sections for parallel ($\parallel$) and perpendicular ($\perp$)
polarizations are given by
\begin{equation}
  \label{eq4}
  \sigma_{\parallel} = 2\rho_{3/2,1/2}\sigma_{3/2} + \rho_{1/2,1/2}\sigma_{1/2},
\end{equation}
\begin{equation}
  \label{eq5}
  \sigma_{\perp} = \frac{1}{2}\{\rho_{3/2,1/2}+3\rho_{3/2,3/2}\}
  \sigma_{3/2} + \rho_{1/2,1/2}\sigma_{1/2}.
\end{equation}
Using {\em ab initio} values for the $m$-averaged cross sections
$\sigma_{3/2}$ ($1s \rightarrow 4p_{3/2}$) and $\sigma_{1/2}$ ($1s
\rightarrow 4p_{1/2}$) \cite{PaBe05}, equations~(\ref{eq4}) and
(\ref{eq5}) may be fitted to the experimental data to determine the
$\rho_{j,\vert m \vert}$. As may be seen in table~\ref{tab1}, the
adiabatic strong-field ionization model we use \cite{SaDu06} reproduces
the general trend displayed by the experimental quantum state
populations.

\section{X-ray absorption by laser-dressed atoms}
\label{sec3}

We now turn our attention to x-ray absorption by noble-gas atoms in a
laser field of intensity $10^{13}$~W/cm$^2$ or lower
\cite{BuSa07,BSYo07}. In this regime, both inner-shell and valence
electrons in the electronic ground state remain essentially unperturbed
(this applies particularly to light atomic species such as neon). The
primary effect of the laser field on x-ray absorption is to modify
(``dress'') the inner-shell-excited electronic states in the vicinity of
an inner-shell edge. Therefore, this laser-dressing approach to
strong-field control of x-ray absorption is reversible and leaves the
target intact. We assume in this section that the laser and x-ray
polarizations are parallel.

The reason why high intensities are required in order to observe
laser-dressing effects in the x-ray absorption spectrum is easy to
understand. The Rabi frequencies associated with laser coupling of the
inner-shell-excited states must be larger than the decay widths of these
states. Inner-shell-excited atoms, which relax via Auger decay and/or
x-ray fluorescence, decay at least a thousand times faster than
valence-excited atoms. Thus, since the dipole coupling matrix elements
between Rydberg levels are similar in the valence- and
inner-shell-excited cases, the laser intensity must be at least $10^6$
times higher than what is needed when all electromagnetic fields
involved are in the optical regime.

\begin{figure}[ht]
  \begin{minipage}{15pc}
    \includegraphics[clip,width=15pc]{fig2.eps}
    \caption{\label{fig2} Calculated dependence of the $1s \rightarrow 3p$ x-ray
         absorption cross section of Ne on the intensity of
         the dressing laser.
         The laser and x-ray polarizations are assumed to be parallel.}
  \end{minipage}\hspace{2pc}%
  \begin{minipage}{20pc}
    \includegraphics[width=20pc]{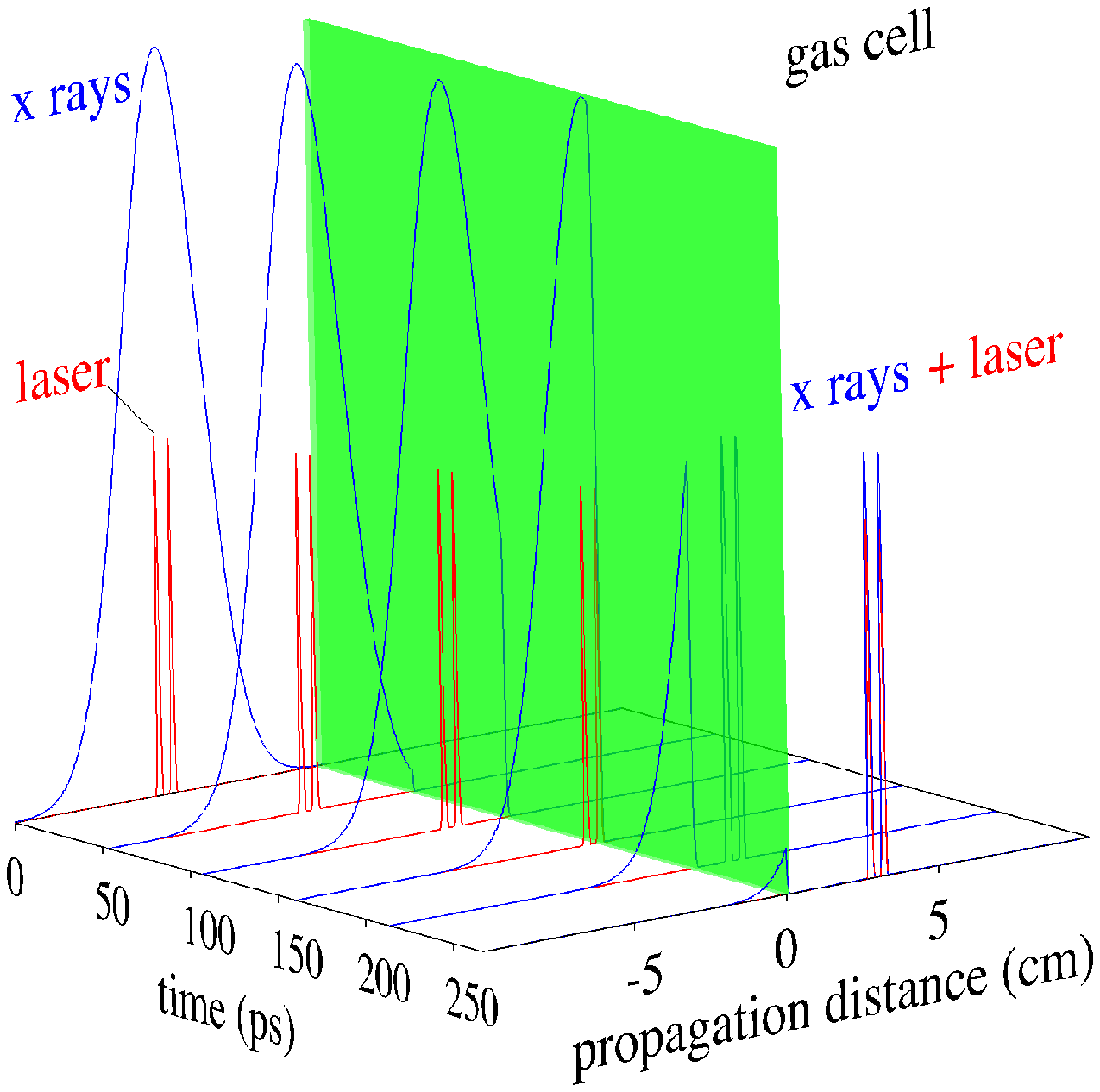}
    \caption{\label{fig3} Generation of ultrafast x-ray pulses using laser
         dressing of Ne.
         See the text for details.}
  \end{minipage}
\end{figure}

An {\em ab initio} theory for calculating the x-ray absorption spectrum
of an atom exposed to an intense laser field is developed in
Ref.~\cite{BuSa07}. The method utilizes the Hartree-Fock-Slater
approximation \cite{MaCo68} to describe the electronic structure of the
unperturbed atom. The coupling to the laser is treated nonperturbatively
within the strong-field (Floquet \cite{Shir65}) limit of nonrelativistic
quantum electrodynamics. The laser-dressed, inner-shell-excited states
are calculated by diagonalizing a non-Hermitian Hamiltonian matrix. The
x-ray-induced transition from the atomic ground state to the
laser-dressed, inner-shell-excited states is treated using perturbation
theory. At a laser intensity of $10^{13}$~W/cm$^2$, the application of
this theory to Kr near the {\em K} edge at $14.3$~keV demonstrates a
$\sim 20\%$ suppression of x-ray absorption at the $1s \rightarrow 5p$
transition \cite{BuSa07}. The suppression is mainly due to laser
coupling of $5p$ to $5s$ and $4d$. In view of the $2.7$-eV decay width
of {\em K}-shell-excited Kr \cite{ChCr80}, the laser intensity it would
take to induce a much stronger modification of the near-{\em K}-edge
structure of Kr is so high that Kr in its ground state would be
strong-field-ionized (cf. section \ref{sec2}).

In neon, the laser-dressing effect is much more dramatic \cite{BSYo07}.
This is because the decay width of {\em K}-shell-excited Ne
\cite{Schm97} is smaller than in Kr by a factor of $10$. The laser
intensity dependence of the x-ray absorption cross section of Ne at the
prominent $1s \rightarrow 3p$ transition ($867$ eV) is displayed in
figure~\ref{fig2}. Laser dressing at an intensity of $10^{13}$~W/cm$^2$
is found to suppress x-ray absorption at the $1s \rightarrow 3p$
resonance by a factor of $13$. This suppression of x-ray absorption is a
manifestation of electromagnetically induced transparency (EIT)
\cite{HaFi90,BoIm91,FlIm05}. To describe EIT one usually considers three
eigenstates of the atom under consideration. To a first approximation in
our case, these states are the Ne ground state (level $1$), the excited
state of Ne with a {\em K}-shell hole and an electron in the $3p$
orbital (level $2$), and the excited state of Ne with a {\em K}-shell
hole and an electron in the $3s$ orbital (level $3$). The laser creates
a coherent superposition of levels $2$ and $3$, so-called dressed
states. At the x-ray photon energy corresponding to the
laser-unperturbed $1 \rightarrow 2$ transition, the transition pathways
from level $1$ to the two dressed states interfere destructively. Hence,
over a narrow x-ray energy range of $\sim 100$~meV, absorption of the x
rays is suppressed at the photon energy corresponding to the transition
from level $1$ to level $2$. Because of the high laser intensity
required for EIT in the x-ray domain, the three-level model provides at
best a qualitatively correct picture. For instance, at
$10^{13}$~W/cm$^2$, the laser ionizes the Rydberg electron at a rate
that is comparable to the Auger decay rate of the {\em K}-shell hole,
i.e., there is strong laser coupling to states outside the three-level
model space.

The ability to control x-ray absorption in Ne at the $1s \rightarrow 3p$
resonance using a strong laser field allows one to imprint pulse shapes
of the optical dressing laser onto the x rays. The idea is illustrated
in figure~\ref{fig3}. Using a $2$-mm long gas cell filled with one
atmosphere of neon, an x-ray pulse at the $1s \rightarrow 3p$ resonance
energy is practically completely absorbed in the gas cell \cite{BSYo07}.
The typical duration of an x-ray pulse from a third-generation
synchrotron is $\sim 100$~ps. Such an x-ray pulse may be overlapped in
time and space with one or several ultrashort, intense laser pulses.
Those portions of the x-ray pulse that overlap with the laser are
transmitted through the gas cell. In the case shown in
figure~\ref{fig3}, where the two dressing laser pulses have a peak
intensity of $10^{13}$~W/cm$^2$, the intensity of the two transmitted
x-ray pulses is about $60\%$ of the incoming x-ray intensity. The time
delay between the two x-ray pulses can be controlled by changing the
time delay between the two laser pulses, opening a route to ultrafast
all x-ray pump-probe experiments. With an analogous strategy, controlled
shaping of short-wavelength pulses might become a reality. A
disadvantage of the method is that it is applicable only at certain
x-ray energies.

\section{X-ray absorption by laser-aligned molecules}
\label{sec4}

The schemes discussed in sections~\ref{sec2} and \ref{sec3} rely on
modifying the system's electronic structure. In the following, we
present a third scheme for controlling resonant x-ray absorption that is
conceptually somewhat different. To motivate this scheme, recall from
section~\ref{sec2} that Kr$^+$ ions produced in a strong laser field are
aligned along the laser polarization axis. Kr$^+$ ions and Br atoms are
isoelectronic and have essentially the same electronic structure. In
contrast to Kr$^+$, Br forms many molecules. If the open $4p$ shell of
Br is not filled upon molecule formation, and if the effective $4p$ hole
orbital is aligned with respect to the molecular frame, then the hole
orbital may be aligned in space by aligning the molecule.

\begin{figure}[ht]
  \includegraphics[clip,width=25pc]{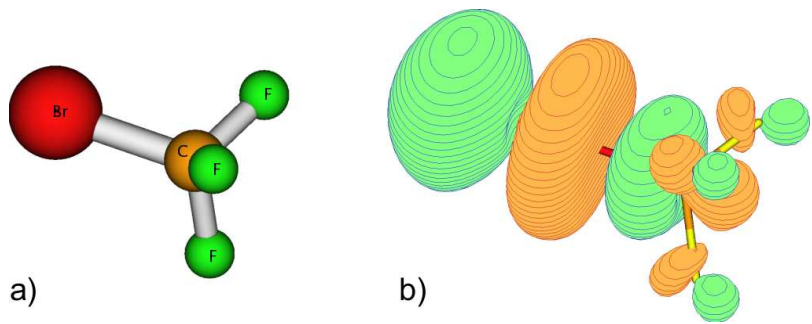}\hspace{2pc}%
  \begin{minipage}[b]{10pc}
    \caption{\label{fig4} a)~Equilibrium geometry of CF$_3$Br in its electronic
         ground state.  b) Lowest unoccupied molecular orbital of CF$_3$Br.}
  \end{minipage}
\end{figure}

The molecule selected for experimental investigations at Argonne's
Advanced Photon Source is CF$_3$Br, which is known as
bromotrifluoromethane or halon-1301 \cite{ClWa94}. The equilibrium
geometry of this molecule in its electronic ground state is shown in
figure~\ref{fig4}a. A Mulliken population analysis \cite{SzOs96} we have
performed on the Hartree-Fock ground-state wave function of CF$_3$Br
indicates that the carbon atom carries a partial charge of $+0.71$,
whereas the partial charge on each of the three fluorine atoms is
$-0.23$. The Br atom acquires essentially no charge (partial charge of
$-0.02$). It may therefore be expected that the Br hole orbital remains
at least partially unoccupied in CF$_3$Br. This is consistent with the
fact that the lowest unoccupied molecular orbital of CF$_3$Br, which is
displayed in figure~\ref{fig4}b, is a $\sigma^{\ast}$ orbital with
substantial Br $4p_z$ character. Here, $z$ refers to the C-Br axis. The
highest occupied molecular orbital of CF$_3$Br is a doubly degenerate
$\pi$ orbital corresponding to Br $4p_x$ and $4p_y$. Owing to the strong
localization of the Br $1s$ orbital near the Br nucleus, an
x-ray-induced transition from the Br $K$ shell to the $\sigma^{\ast}$
orbital probes only the atomic $p_z$ character of the $\sigma^{\ast}$
orbital. Let $D_x$, $D_y$ and $D_z$ denote the Cartesian components,
with respect to the molecular frame, of the transition dipole vector
between Br $1s$ and $\sigma^{\ast}$. The molecule CF$_3$Br belongs to
the $C_{3\mathrm{v}}$ point group. Hence, $D_x=D_y=0$, for Br $1s$ and
the $\sigma^{\ast}$ orbital span the irreducible representation $A_1$.
Only when the x-ray polarization vector has a component along the C-Br
axis can x-ray absorption at the Br $1s \rightarrow \sigma^{\ast}$
resonance take place. Estimates show that the intramolecular electric
field in CF$_3$Br breaks spin-orbit coupling in Br, so that in contrast
to the situation we discussed in connection with Kr$^+$, spin-orbit
coupling may be neglected.

The polarization dependence of resonant x-ray absorption by aligned
molecules was demonstrated in studies on surface-adsorbed molecules
\cite{HaCe00} and fixed-in-space molecules \cite{AdKo05}. This effect
may be exploited for a laser-based control scheme for x-ray absorption.
As explained in reference~\cite{StSe03} and references therein, small
molecules with an anisotropic polarizability may be aligned along the
laser polarization axis at laser intensities in the range $10^{11}$ to
$10^{12}$~W/cm$^2$. The physics underlying laser-induced molecular
alignment is easy to understand: The instantaneous laser electric field
polarizes the molecule. The electric dipole moment induced depends on
the molecular polarizability tensor and the laser electric field vector
with respect to the molecular frame. The interaction of the induced
dipole moment with the laser electric field leads to an effective
potential for the rotational degrees of freedom of the molecule. The
effective laser-induced potential is proportional to the cycle-averaged
laser intensity. The potential-energy minimum is reached when the most
polarizable molecular axis is parallel to the laser polarization axis.
In the case of a symmetric rotor, the molecule-specific quantity that
characterizes the effective potential is $\alpha_{\parallel} -
\alpha_{\perp}$, which is the difference between the polarizability
component of the molecule along the molecular symmetry axis and the
polarizability component perpendicular to the molecular symmetry axis.
An assumption typically made is that the laser photon energy is far too
low to electronically excite the molecule via a one-photon transition.
Consequently, it is permissible to use the {\em static} polarizability
components obtained in the limit of infinite wavelength. We calculated
$\alpha_{\parallel} - \alpha_{\perp}$ using coupled-cluster linear
response theory \cite{Dalt05} and found that for CF$_3$Br the difference
between $\alpha_{\parallel} - \alpha_{\perp}$ determined in the static
limit and $\alpha_{\parallel} - \alpha_{\perp}$ determined at $800$~nm
is indeed only about $2\%$.

For the purpose of minimizing temporal averaging effects, it is
important that the duration of the molecular alignment is not
substantially shorter than the x-ray pulse duration of $\sim 100$~ps.
This means that the impulsive alignment technique \cite{StSe03}, where
alignment persists for only a few ps, is unsuitable. In the adiabatic
regime \cite{StSe03}, where the laser intensity envelope changes slowly
on the characteristic rotational time scale of the molecule, the
alignment persists for as long as the laser pulse is on. CF$_3$Br is a
prolate symmetric top molecule \cite{Krot03} with a rotational period of
$239$~ps. (This is the characteristic time for rotation about an axis
perpendicular to the C-Br axis.) In order to obtain a high laser
repetition rate ($1$~kHz), the laser pulse energies were limited to
about $2$~mJ with a pulse duration of $95$~ps. Even though this is
somewhat shorter than the rotational period, we find in our
calculations, which are similar to the ones reported in
reference~\cite{HaSe05}, that at temperatures of several K or higher,
the molecular response to the laser pulse is essentially adiabatic and
tracks the laser pulse envelope.

\begin{figure}
  \begin{center}
    \includegraphics[clip,width=\hsize]{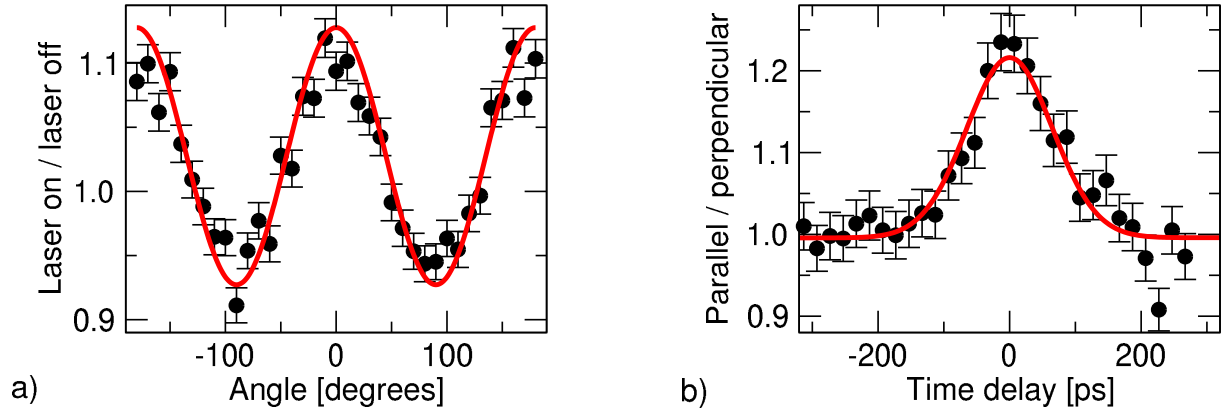}
    \caption{Resonant x-ray absorption by strong-field-aligned CF$_3$Br.
         Comparison between theory (solid line) and experiment (circles
         plus error bars).
         a)~Dependence of the x-ray absorption signal on the angle between
         the laser and x-ray polarization axes.
         b)~Ratio between x-ray absorption signal for parallel laser and
         x-ray polarizations and x-ray absorption signal for perpendicular
         laser and x-ray polarizations, as a function of the time delay
         between the laser and x-ray pulses.}
    \label{fig5}
  \end{center}
\end{figure}

Figure~\ref{fig5} shows a comparison between experimental data taken
recently at the Advanced Photon Source and a theory we have developed
for this purpose. The peak laser intensity was $0.85\times
10^{12}$~W/cm$^2$. The experimental setup was similar to the Kr
experiment mentioned in section~\ref{sec2}. {\em K}$\alpha$ fluorescence
was used to monitor $1s$-hole formation in Br. The Br near-{\em K}-edge
absorption spectrum was measured and the $1s \rightarrow \sigma^{\ast}$
resonance at $13.476$~keV was identified. The experimental data plotted
in figure~\ref{fig5} were obtained on resonance, after subtraction of a
$10\%$ background from inner-shell-excited Rydberg and continuum states
that overlap with the resonance. The primary experimental difference
from the atomic Kr experiment is that a supersonic expansion of $5\%$
CF$_3$Br seeded in He was used to produce a cold jet of molecules. By
applying a high backing pressure to the nozzle through which the gas
mixture expands, a low rotational temperature may be achieved. This is
crucial for alignment in the adiabatic regime \cite{StSe03}. In the
experiment at the Advanced Photon Source, a backing pressure of up to
$9$ bar was applied, which by comparison with reference~\cite{KuBi06}
suggests a rotational temperature of the order of $10$~K.

The central quantities we calculate for comparison with experiment are
on-resonance absorption cross sections for parallel laser and x-ray
polarizations ($\sigma_{\parallel}$); perpendicular laser and x-ray
polarizations ($\sigma_{\perp}$); and a laser-free thermal ensemble
($\sigma_{\mathrm{th}}$). Let $J\I{X}(t)$ denote the x-ray photon
current density (pulse envelope). The x-ray absorption probability per
pulse is then given by
\begin{equation}
  P_i(\tau) = \Int_{-\infty}^{\infty} \sigma_i(t) \, J\I{X}(t-\tau) \differential t, \qquad i=\parallel,\perp,\mathrm{th},
\end{equation}
where $\tau$ is the time delay between the laser and x-ray pulses, which
are both assumed to be Gaussian. At $\tau=0$, the maxima of the laser
and x-ray pulses coincide. One quantity that may be measured is the
dependence of the x-ray absorption on the angle $\vartheta\I{LX}$
between the laser and x-ray polarizations. This is shown in
figure~\ref{fig5}a ($\tau=0$), where the ratio between the laser-on and
laser-off signals is plotted. The theoretical expression for the
$\vartheta\I{LX}$ dependence of the laser-on/laser-off ratio is
\begin{equation}
  R(\vartheta\I{LX}) = \frac{P_{\parallel}(0)}{P_{\mathrm{th}}(0)} \, \cos^2
\vartheta\I{LX} +
            \frac{P_{\perp}(0)}{P_{\mathrm{th}}(0)} \, \sin^2 \vartheta\I{LX}.
\end{equation}
Assuming a rotational temperature of $21$~K and an x-ray pulse duration
of $127$~ps, good agreement is found between experiment and theory.
Note, in particular, that the theory correctly predicts that
$|R(0^{\circ}) - 1| > |R(90^{\circ})-1|$. Figure~\ref{fig5}b shows the
ratio $P_{\parallel}(\tau)/P_{\perp}(\tau)$ as a function of the time
delay $\tau$. Agreement between theory and experiment is again
satisfactory for a temperature of $21$~K and an x-ray pulse duration of
$127$~ps.

In conclusion, we would like to mention some future challenges and
opportunities. We discussed in section~\ref{sec2} a spectroscopic probe
of strong-field-ionized noble-gas atoms. The {\em coherence} properties
of the ion density matrix produced in a strong optical field are not
known and pose a challenge for future experiments with ultrafast
radiation pulses. The EIT effect for x rays explored in
section~\ref{sec3} may make it possible to perform ultrafast x-ray pulse
shaping. Such pulses could have useful applications in time-resolved
x-ray science. Finally, in this section we demonstrated that
laser-induced alignment allows one to control resonant x-ray absorption
in molecules. Calculations indicate that by reducing the molecular
rotational temperature to a few K, and by using an x-ray pulse that is
shorter than the aligning laser pulse, it should be possible to suppress
x-ray absorption for perpendicular laser and x-ray polarizations by a
factor of five or so relative to the parallel case.

\ack
We would like to thank D~A~Arms, D~L~Ederer, E~C~Landahl and S~T~Pratt
for their assistance with some of the experiments. This work was
supported by the Office of Basic Energy Sciences, Office of Science,
U.S. Department of Energy, under Contract No.~DE-AC02-06CH11357. C~Buth
was partly supported by the Alexander von Humboldt Foundation.

\end{document}